\documentclass[showpacs,aps,prl,reprint,superscriptaddress,preprintnumbers]{revtex4-1}

\usepackage{amsmath}
\usepackage{amssymb}
\usepackage{graphicx}
\usepackage{dcolumn}
\usepackage[dvipsnames]{xcolor}
\usepackage{amsthm,amssymb}
\usepackage{mathrsfs}
\usepackage{color}
\usepackage{endnotes}
\usepackage{bm}
\usepackage{epsfig}
\usepackage{array}
\usepackage{ulem}
\usepackage{color,soul}

\newcommand{\etal}{{\it et al.}}
\newcommand{\ie}{{\it i.e.\ }}
\newcommand{\CsVSb}{{CsV$_{3}$Sb$_{5}$}}
\newcommand{\AVSb}{{AV$_{3}$Sb$_{5}$}} 
\newcommand{\CaRhSn}{(Ca$_{x}$Sr$_{1-x}$)$_{3}$Rh$_{4}$Sn$_{13}$}

\begin{document}

\title{Quantum Phase Transition as a Promising Route\\ to Enhance the Critical Current in Kagome Superconductor CsV$_{3}$Sb$_{5}$}

\author{Wenyan~Wang}
\author{Lingfei~Wang}
\author{Xinyou~Liu}
\author{Chun~Wai~Tsang}
\author{Zheyu~Wang}
\author{Tsz~Fung~Poon}
\affiliation{Department of Physics, The Chinese University of Hong Kong, Shatin, Hong Kong, China}
\author{Shanmin~Wang}
\affiliation{Department of Physics, Southern University of Science and Technology, Shenzhen, Guangdong, China}
\author{Kwing~To~Lai}
\author{Wei~Zhang}
\affiliation{Department of Physics, The Chinese University of Hong Kong, Shatin, Hong Kong, China}
\author{Jeffery~L.~Tallon}
\affiliation{Robinson Institute, Victoria University of Wellington, Wellington 6140, New Zealand}
\author{Swee~K.~Goh}
\email[]{skgoh@cuhk.edu.hk}
\affiliation{Department of Physics, The Chinese University of Hong Kong, Shatin, Hong Kong, China}

\date{October 16, 2024}

\begin{abstract}
Developing strategies to systematically increase the critical current, the threshold current below which the superconductivity exists, is an important goal of materials science. Here, the concept of quantum phase transition is employed to enhance the critical current of a kagome superconductor CsV$_3$Sb$_5$,
which exhibits a charge density wave (CDW) and superconductivity that are both affected by hydrostatic pressure. As the CDW phase is rapidly suppressed under pressure, a large enhancement in the self-field critical current ($I_{\rm c,sf}$) is recorded. The observation of a peak-like enhancement of $I_{\rm c,sf}$ at the zero-temperature limit ($I_{\rm c,sf}(0)$) centred at $p^*\approx 20$~kbar, the same pressure where the CDW phase transition vanishes, further provides strong evidence of a zero-temperature quantum anomaly in this class of pressure-tuned superconductor. Such a peak in $I_{\rm c,sf}(0)$ resembles the findings in other well-established quantum-critical superconductors, hinting at the presence of enhanced quantum fluctuations associated with the CDW phase in CsV$_3$Sb$_5$.

\end{abstract}

\maketitle

\noindent {\bf \large 1. Introduction}\\
\noindent Quantum phase transition, a phase transition at absolute zero temperature triggered by a non-thermal tuning parameter, can stabilize exotic phases such as superconductivity~\cite{shibauchi2014,gegenwart2008}. Across a wide variety of material systems, a dome-shaped superconducting phase is often observed when a neighbouring phase is suppressed by non-thermal parameters such as pressure, strain or chemical doping. Interestingly, the maximum superconducting transition temperature $T_{\rm c}$ is often found near the zero-temperature phase boundary, suggesting that the robustness of the condensate, which is essential for practical applications, is inherently linked to the physics associated with the quantum phase transition. 
In addition to $T_{\rm c}$, the current-carrying capability of a superconductor is fundamentally important and the critical current ($I_{\rm c}$) is another parameter to optimize. Thus, it is crucial to explore if quantum phase transition can also play a positive role in enhancing $I_{\rm c}$.

The recently discovered family of kagome superconductors \AVSb\ (A=K, Rb, Cs) undergoes a charge density wave (CDW) transition at around 100~K, and a superconducting transition below 10~K~\cite{Ortiz2019,Ortiz2020,Du2021,Chen2021a,Li2021,Liang2021,Zhao2021,Liu2021,Zhou2021,Jiang2021,Yu2021b,Yu2021,wang2021,Yang2020,hu2022,Uykur2022,Kang2022a,Wu2022,Lou2022,yu2022,han2023,Kang2023}. Particularly in \CsVSb, it has been pointed out that the hybridization between V and Sb orbitals plays a critical role in mediating the CDW phase transition~\cite{han2023}. The CDW phase itself is complicated, with different charge order patterns reported~\cite{Kang2023,Tan2021,luo2022,Ortiz2021}. Under pressure, the CDW can be suppressed and, for \CsVSb, $T_{\rm c}$ shows a peculiar double-dome behaviour~\cite{Chen2021a,Yu2021}. Interestingly, the second dome in \CsVSb\ is centered around the extrapolated critical pressure $p^*$ where the CDW disappears. Thus, the temperature-pressure ($T$-$p$) phase diagram of \CsVSb\ resembles the phase diagram of many quantum-critical superconductors. The possibility of a quantum critical point (QCP) at $p^*$ has been put forward theoretically~\cite{tazai2022,wang2022}.  On the experimental front, high-pressure nuclear quadrupole resonance reported a diverging resonance linewidth at $p^*$~\cite{feng2023}, and the initial slope of the upper critical field, $-\left(dH_{\rm c2}/dT\right)_{T_{\rm c}}$, was found to peak at $p^*$~\cite{Chen2021a}. Recently, high-pressure Shubnikov-de Haas oscillations detected a noticeable enhancement of the quasiparticle effective masses on approaching $p^*$~\cite{zhang2024}. All these observations suggest the presence of quantum fluctuations that can mediate superconducting pairing. Hence, \CsVSb\ is an ideal system to investigate if quantum phase transition can drive the enhancement of $I_{\rm c}$.

In this work, we measure the self-field transport critical current ($I_{\rm c,sf}$), \ie\ transport critical current without applying an external magnetic field, over a wide pressure range straddling across $p^*$. We observe a drastic enhancement of the critical current when the {\it same} set of samples was tuned by pressure. The use of the same samples eliminates extrinsic factors such as the uncertainty in the geometric factor or random potentials due to impurities. In particular, the zero temperature limit of $I_{\rm c,sf}$, denoted as $I_{\rm c,sf}(0)$, of \CsVSb\ peaks sharply at $p^*$. This observation further confirms the occurrence of a quantum phase transition beneath the superconducting dome. Together with existing experimental data from other probes, our work establishes access to a quantum phase transition as a promising route to optimize $I_{\rm c}$. Finally, we show that the measurement of $I_{\rm c,sf}$ provides a systematic means to extract key superconducting parameters of \CsVSb\ under pressure.
\\

\begin{figure}[!t]\centering
      \resizebox{8.5cm}{!}{
              \includegraphics{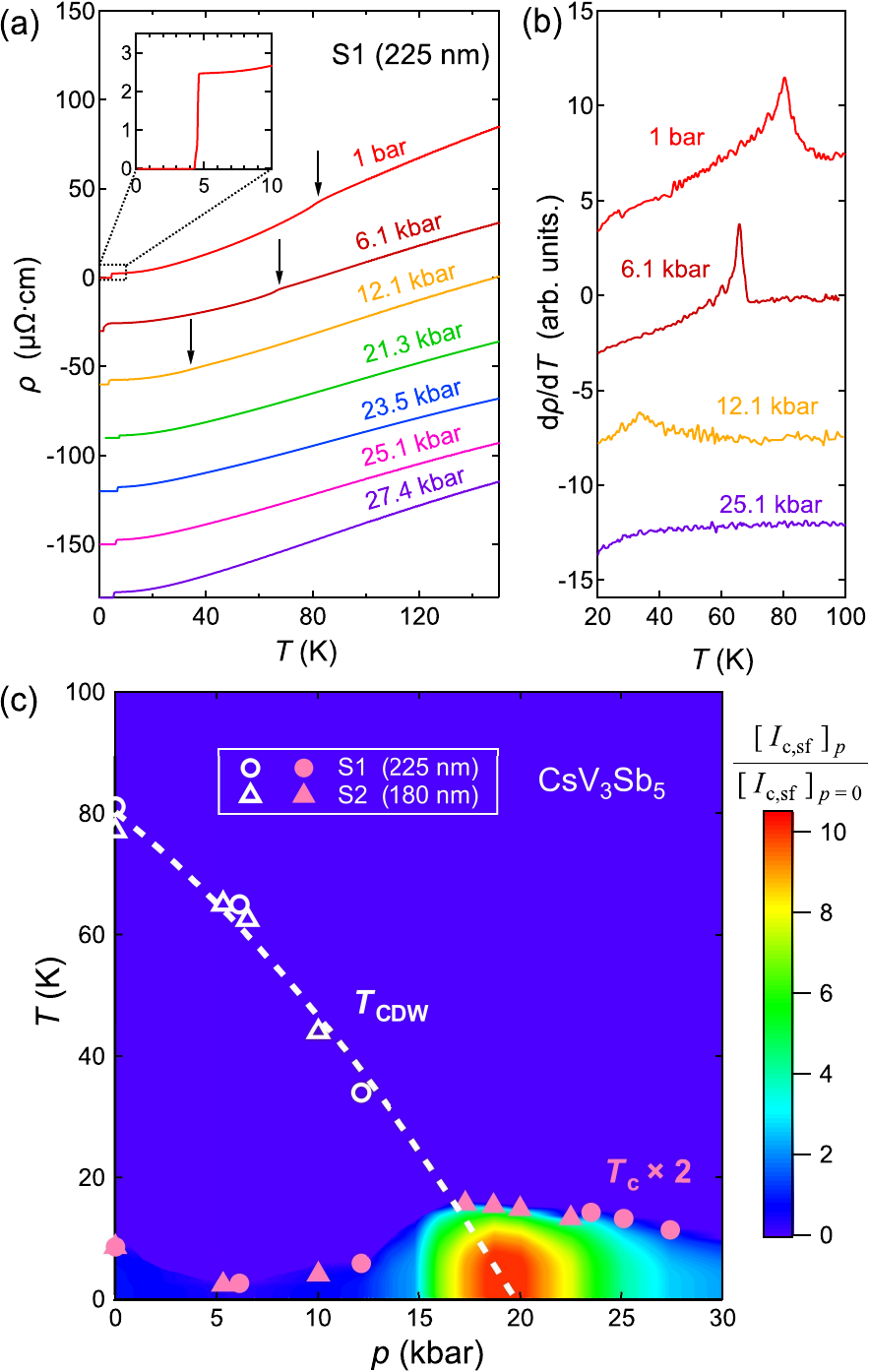}}              				
              \caption{\label{fig1}   
              (a) Temperature dependence of the electrical resistivity $\rho$ at various pressures in \CsVSb\ (S1). All of the traces are measured upon warming up. For clarity, all traces, except the dataset at 1 bar, are successively down-shifted by 30 $\mu\Omega\cdot$cm. The black arrows indicate the $T_{\rm CDW}$ and the inset displays the data near the superconducting transition. (b) Temperature dependence of $d\rho/dT$ in S1 at four typical pressures. $T_{\rm CDW}$ is determined as the temperature at which $d\rho/dT$ peaks. (c) $T$-$p$ phase diagram of \CsVSb\ thin flakes with a contour map of the normalized $I_{\rm c,sf}$ for S1 and S2 overlaid. See the main text for details. White circles and triangles are $T_{\rm CDW}$ of S1 and S2 at various pressures. The white dashed curve indicates the implied, and extrapolated border of CDW phase. The CDW phase is gradually suppressed under pressure and ultimately vanishes at $p^* \approx 20$~kbar. Pink circles and triangles are $T_{\rm c}$ of S1 and S2 at various pressures, demonstrating a double-dome pressure dependence. $T_c$ is determined by the temperature at which the resistivity reaches zero.}   
\end{figure}

\noindent {\bf \large 2. Results and Discussion}\\
\noindent {\bf 2.1 Construction of $T$-$p$ Phase Diagram}\\
\noindent \textbf{Figure~\ref{fig1}}a displays the temperature dependence of resistance for \CsVSb\ (S1), which has a thickness of 225~nm, at various pressures. At ambient pressure, a kink is visible at $T_{\rm CDW}=80$~K, corresponding to the CDW transition well-documented in the literature for thin \CsVSb\ flakes~\cite{song2021,zhang2023,zhang2022,zheng2023,song2023,ye2024}. The transport signature associated with the CDW transition can be best visualized in the thermal derivative $d\rho/dT$ (Figure~\ref{fig1}b). At a lower temperature, the electrical resistivity vanishes at $T_{\rm c}=4.3$~K (see inset of Figure~\ref{fig1}a), signifying a superconducting transition, consistent with previous reports on thin flakes of \CsVSb~\cite{song2021,zhang2023,ye2024}. The higher $T_{\rm c}$ in \CsVSb\ thin flakes could be attributed to the biaxial strain effect, which is unavoidable in device architecture (Figure S5, Supporting Information).

Under pressure, the anomaly associated with the CDW transition is weakened. To track $T_{\rm CDW}$, we follow the peak in $d\rho/dT$. At 12.1~kbar, $T_{\rm CDW}$ decreases to 34~K. Concomitantly, $T_{\rm c}$ first decreases under pressure then begins to increase. The data for another flake (\CsVSb~(S2)) show a similar pressure dependence. Combining all data from S1 and S2, a $T$-$p$ phase diagram can be constructed (Figure~\ref{fig1}c), which clearly shows that $T_{\rm c}$ peaks around the pressure $p^*$ where 
$T_{\rm CDW}$ extrapolates to 0~K. Unlike the $T$-$p$ phase diagram of the bulk sample, the maximum of the lower-pressure dome shifts from $\sim$7~kbar to or even below ambient pressure in our flake, while the high-pressure dome remains centered at $\sim$20~kbar. Our $T$-$p$ phase diagram is similar to that constructed by Ye \etal~\cite{ye2024}. As a preview, we overlay the temperature dependence of $I_{\rm c,sf}$ at various pressures as a contour map on the $T$-$p$ phase diagram. We can see that $I_{\rm c,sf}$ also experiences a drastic enhancement near $p^*$. In the remaining texts, we will demonstrate the construction of this contour map.
\\

\begin{figure}[!t]\centering
      \resizebox{8.5cm}{!}{
 \includegraphics{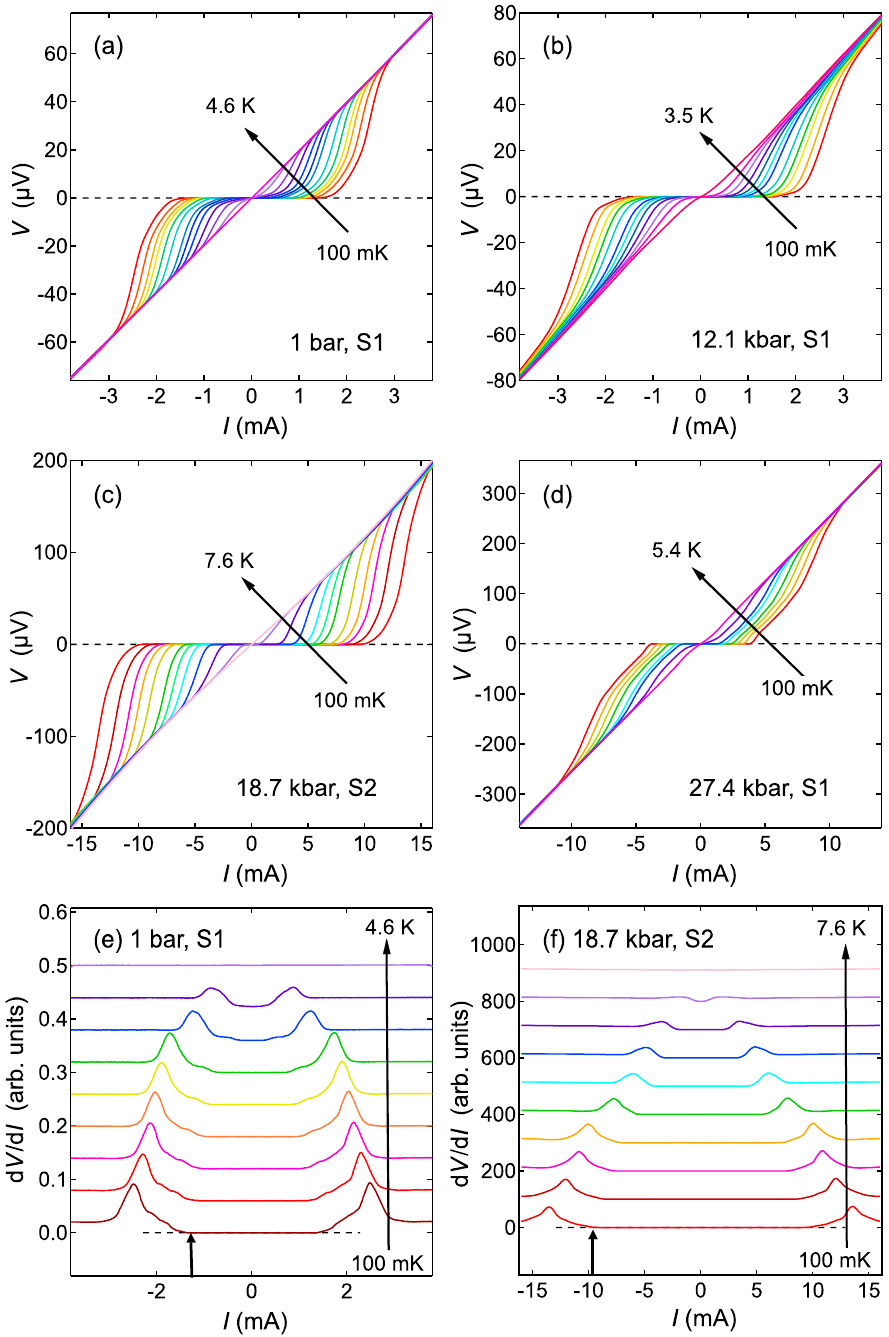}}       
 \caption{\label{fig2} Temperature dependence of $V$-$I$ relations at (a) ambient pressure, (b) 12.1 kbar, (c) 18.7 kbar and (d) 27.4 kbar in S1 (225~nm) and S2 (180~nm). The solid arrows indicate the increase of temperature from 100~mK to just above $T_{\rm c}$. The calculated first derivative of $V(I)$, $dV/dI$ of (e) S1 at ambient pressure and (f) S2 at 18.7 kbar. The short arrows indicate the value of self-field critical current, which is defined when $dV/dI$ first deviates from zero. The long arrows demonstrate the increase of temperature from 100~mK to the temperature just above $T_{\rm c}$. All $V$-$I$ curves and their derivatives are shown in Figure S2 and Figure S3 (Supporting Information).}
          
\end{figure}

\noindent {\bf 2.2 $I_{\rm c,sf}$ as a Probe of the Superconducting Gap}\\
Based on the $T$-$p$ phase diagram constructed, we measure a series of voltage-current ($V$-$I$) curves in the superconducting state at various pressures. Representative data are shown in \textbf{Figure~\ref{fig2}}a-d. Above $T_{\rm c}$ ({\it e.g.} 4.6~K at 1 bar), $V$ is linear in $I$, showing a typical ohmic behaviour of a metal. At the lowest temperature for each pressure ({\it e.g.} 100~mK at 1 bar), $V$ is zero for low current but increases drastically when the current reaches a threshold. This is a typical critical current behaviour. The $V$-$I$ curves for the intermediate temperatures evolve systematically between the two limits.

Following the same practice as our earlier work~\cite{zhang2023}, we define $I_{\rm c,sf}$ as the current at which $dV/dI$ deviates from zero.  Representative $dV/dI$ curves are displayed in Figure~\ref{fig2}e,f, with the short vertical arrows denoting $I_{\rm c,sf}$ at 100~mK for 1 bar and 18.7~kbar, respectively. The $dV/dI$ curves clearly demonstrate the smooth evolution of $I_{\rm c,sf}$ as a function of temperature. We also note the different current scales in Figure~\ref{fig2}e,f. Therefore, $I_{\rm c,sf}$ is roughly an order of magnitude larger at 18.7~kbar than that at 1~bar. 

\begin{figure}[!t]\centering
       \resizebox{8.5cm}{!}{
              \includegraphics{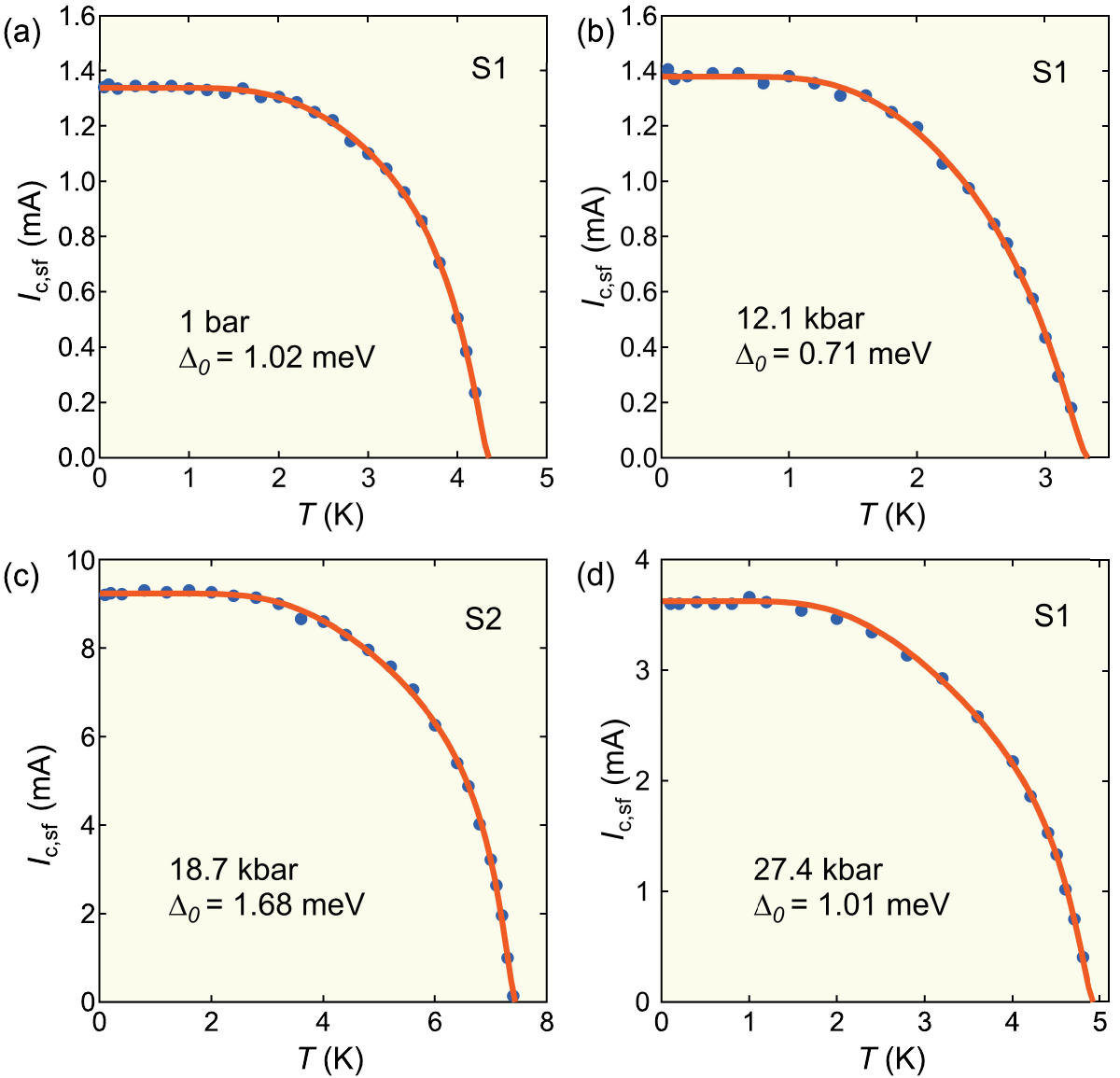}}                			
              \caption{\label{fig3} Temperature dependence of $I_{\rm c,sf}$ at four representative pressures. The solid curves are the single $s$-wave gap fits. The extracted superconducting gap $\Delta_0$ are 1.02 meV at ambient pressure, 0.71 meV at 12.1 kbar, 1.68 meV at 18.7 kbar and 1.01 meV at 27.4 kbar, respectively. $I_{\rm c,sf}(T)$ at all pressures are shown in Figure S4 (Supporting Information).
}
\end{figure}

To have a detailed examination of the pressure dependence of $I_{\rm c,sf}(T)$, the extracted $I_{\rm c,sf}(T)$ from Figure~\ref{fig2} are plotted in \textbf{Figure~\ref{fig3}} at various pressures. 
When the temperature decreases from $T_{\rm c}$, $I_{\rm c,sf}(T)$ at first rises rapidly and then saturates when the temperature is sufficiently low. $I_{\rm c,sf}(T)$ curves at all pressures are qualitatively similar. Quantitative information can be obtained by noting the physical significance of $I_{\rm c,sf}$. A series of studies~\cite{talantsev2015,talantsev2017,zhang2023,liu2022,semenok2022} have shown that the transport critical current density at zero magnetic field, $J_{\rm c,sf}$, is related to the superfluid density $\rho_{\rm s} \propto \lambda^{-2}$, where $\lambda$ denotes the penetration depth. Specifically, when the half thickness $b$ of the flake is less than $\lambda$, $J_{\rm c,sf}$ is given by~\cite{talantsev2015,talantsev2017}
\begin{equation}
J_{\rm c,sf}=\frac{\phi_0}{4\pi\mu_0\lambda^3}\left(\ln\left(\frac{\lambda}{\xi}\right)+0.5\right),
\label{eqn_thin}
\end{equation}
where $\phi_0$ is the flux quantum, $\mu_0$ is the vacuum permeability and $\xi$ is the coherence length. 

The thickness $2b$ of S1 and S2 are 225~nm and 180~nm, respectively. Furthermore, the penetration depth at the 0~K limit, $\lambda(0)$, is approximately 450 nm~\cite{gupta2022microscopic}. Thus, the thin-limit Equation~(\ref{eqn_thin}) is appropriate here. The temperature dependence of $J_{\rm c,sf}$ is dominated by the temperature dependence of $\lambda$, with a reasonable assumption that the temperature dependence of $\lambda/\xi$ under the logarithm can be disregarded. Consequently, $J_{\rm c,sf}(T)$/$J_{\rm c,sf}(0)$ directly reflects~$[\lambda^{-2}(T)/\lambda^{-2}(0)]^{3/2}$, which is influenced by the magnitude of the superconducting gap.  From the NMR measurements, a Hebel-Slichter coherence peak just below $T_{\rm c}$ suggests \CsVSb\ exhibits an $s$-wave superconductivity~\cite{mu2021s}. Moreover, both the muon spin rotation ($\mu$SR) and the tunnel diode oscillator (TDO) results of magnetic penetration depth indicate that the superconducting order parameter has a two-gap $s$-wave symmetry, indicating nodeless superconductivity in \CsVSb~\cite{gupta2022microscopic,gupta2022two,shan2022muon,duan2021nodeless}. Furthermore, angle-resolved photoemission spectroscopy measurements on \CsVSb, tuned by isovalent Nb/Ta substitutions of V, also suggest a nodeless electron pairing~\cite{zhong2023nodeless}. Finally, the previous self-field critical current study by some of us also demonstrates nodeless superconductivity in \CsVSb, regardless of the presence of time reversal symmetry~\cite{zhang2023}. Therefore, we attempt to fit the low-temperature $J_{\rm c,sf}(T)$ by taking an $s$-wave gap, where $\lambda^{-2}(T)/\lambda^{-2}(0)$ is given by~\cite{muhlschlegel1959}:
\begin{equation}
\frac{\rho_{\rm s}(T)}{\rho_{\rm s}(0)}=\frac{\lambda^{-2}(T)}{\lambda^{-2}(0)}=1-2\sqrt{\frac{\pi\Delta_0}{k_{\rm B}T}}\exp\left(-\frac{\Delta_0}{k_{\rm B}T}\right).
\label{swave}
\end{equation}

\begin{figure}[!t]\centering
      \resizebox{8.5cm}{!}{
 \includegraphics{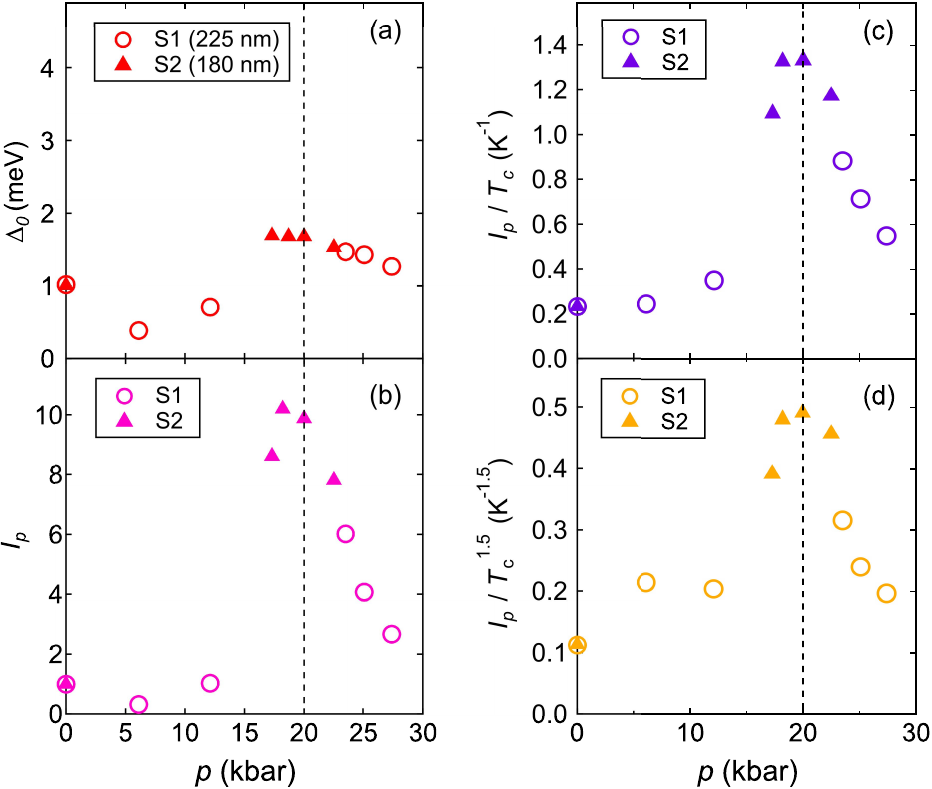}}       
 \caption{\label{fig4} Pressure dependence of (a) the superconducting gap $\Delta_0$, (b) $I_{p} = [I_{\rm c,sf}(0)]_p/[I_{\rm c,sf}(0)]_{p=0}$, (c) $I_{p}/T_{\rm c}$, (d) $I_{p}/T_{\rm c}^{1.5}$. The open circles and solid triangles in (a)-(d) denote S1 and S2, respectively. Both S1 and S2 have a thickness around 200~nm. In (a)-(d), $p^*$ is denoted by the vertical dashed line. 
}            
\end{figure}

As shown in Figure~\ref{fig3}, the combination of Equation~(\ref{eqn_thin}) and Equation~(\ref{swave}) can describe the experimental data very well,  even at higher temperatures. From these analyses, the pressure dependence of the superconducting gap $\Delta_0$ can be extracted. At ambient pressure, the superconducting gap $\Delta_0$ = 1.02~meV (2.75~$k_{\rm B}T_{\rm c}$), which is larger than the BCS weak coupling limit and consistent with the previous results 0.95 meV (2.70~$k_{\rm B}T_{\rm c}$) and 1.08 meV (2.84~$k_{\rm B}T_{\rm c}$) ~\cite{zhang2023}. At higher pressures, $\Delta_0$ varies in a non-monotonic manner (\textbf{Figure~\ref{fig4}}a) but appears to track the pressure dependence of $T_{\rm c}$. However, the gap-to-$T_{\rm c}$ ratio shows a noticeable pressure dependence -- $2\Delta_0/k_{\rm B}T_{\rm c}$ ranges from 4.8 to 6.9 throughout the entire pressure range. 
\\
\\
\noindent {\bf 2.3 Enhancement of $I_{\rm c,sf}$ on Approaching $p^*$}\\
We now address the most notable feature of our dataset, namely the drastic pressure dependence of $I_{\rm c,sf}$. Using the fitted $I_{\rm c,sf}$ extrapolated to 0~K as $I_{\rm c,sf}(0)$, we plot its pressure dependence in Figure~\ref{fig4}b. To facilitate the comparison of different samples, we normalize $I_{\rm c,sf}(0)$ under pressure to the ambient pressure value [$I_{\rm c,sf}(0)]_{p=0}$ for each sample. Thus, near $p^*$, $I_{p} = [I_{\rm c,sf}(0)]_p/[I_{\rm c,sf}(0)]_{p=0}$ increases by nearly 10-fold compared with the value at ambient pressure.
To address whether the enhancement of $I_{\rm c,sf}(0)$ is solely due to the increase in $T_{\rm c}$ near $p^*$, we explored the relationship between $I_{\rm c,sf}$ and $T_{\rm c}$. As the simplest scenario, we plot the pressure dependence of $I_p$ divided by $T_{\rm c}$ in Figure~\ref{fig4}c. In the vicinity of $p^*$, the ratio increases by approximately 6-fold. In our thin flakes, we presume $I_{\rm c,sf}(0) \propto J_{\rm c,sf}(0) \propto \lambda^{-3}$, according to Equation~(\ref{eqn_thin}). Assuming the applicability of Uemura relation, which states that $T_{\rm c} \propto \lambda^{-2}$~\cite{uemura1989}, we then have $I_{\rm c,sf}(0) \propto T_{\rm c}^{1.5}$. Thus, we also plot the pressure dependence of $I_{p}/T_{\rm c}^{1.5}$ (Figure~\ref{fig4}d). As can be seen, near $p^*$, $I_{p}/T_{\rm c}^{1.5}$ is still markedly enhanced by approximately 5-fold over the ambient-pressure ratio. Therefore, the enhancement in $I_{\rm c,sf}(0)$ on approaching $p^*$ is genuine, and cannot be merely explained by the $T_{\rm c}$ enhancement. To best visualize the pressure dependence of $I_{\rm c,sf}(T)$ in the context of the phase diagram, a contour map is overlaid as displayed in Figure~\ref{fig1}c: $I_{\rm c,sf}$ is gradually enhanced on approaching $p^* \approx$ 20~kbar.

The peaking of the critical current has been observed and discussed alongside quantum criticality in diverse systems. In heavy fermion CeRhIn$_5$, a peak shows up in the pressure dependence of zero-field $I_{\rm c}$(0) at a hidden quantum critical point (QCP), where a continuous antiferromagnetic transition is suppressed~\cite{jung2018}. Moreover, there is a noticeable peak in $I_{\rm c,sf}(0)$ at a unique doping level $x \approx$ 0.19 in the cuprate Bi$_2$Sr$_2$CaCu$_2$O$_{8+\delta}$~\cite{naamneh2014,talantsev2015}. In another system \CaRhSn, when it is tuned towards the structural QCP by chemical pressure, a significant increase in $J_{\rm c,sf}(0)$ has also been observed~\cite{liu2022}. Hence, based on the aforementioned evidence, the observed peak in $I_{\rm c,sf}(0)$ near $p^*$ in \CsVSb\ thin flake appears to stem from the enhanced quantum fluctuations near the edge of the charge-ordered phase. 

Recently, quantum oscillation studies straddling $p^*$ have detected an enhancement of the quasiparticle effective masses ($m^*$) near $p^*$, highlighting the presence of fluctuations associated with the destabilization of the charge-ordered phase. Because $\lambda^{-2}=\rho_{\rm s}\mu_0e^2/m^*$, the enhancement of $m^*$ near $p^*$ implies a smaller $\lambda^{-2}$ and a suppressed $I_{\rm c,sf}$. To reconcile with our data, the superfluid density $\rho_{\rm s}$ must also increase near $p^*$, and significantly more rapidly than $m^*$. Therefore, it is the increase of the {\it ratio} $(\rho_{\rm s}$/$m^{*})$ that contributes to the peak in critical current.  Indeed, the enhancement of  $(\rho_{\rm s}/m^*)$ at the critical hole concentration 0.19 has long been recognized in cuprates~\cite{tallon2001,bernhard2001,panagopoulos1999,niedermayer1993}, and a similar behaviour can be expected for \CsVSb. Perhaps the most puzzling case concerns BaFe$_2$(As$_{1-x}$P$_x$)$_2$, in which a striking peak in $\lambda^2$ has been observed at a QCP -- the very opposite of what is reported here~\cite{hashimoto2012}. However, $J_{\rm c}$ measured by either magnetic hysteresis loop method~\cite{ishida2017,ishida2018} or by the transport method~\cite{hiramatsu2017} also demonstrates a maximum value when the system is tuned towards the QCP. Next, also for BaFe$_2$(As$_{1-x}$P$_x$)$_2$, the lower critical field, which is  proportional to $\lambda^{-2}$, also peaks at the QCP. Returning to the present case of \CsVSb, in a recent nuclear quadrupole resonance (NQR) measurement, the linewidth $\delta\nu$ of $^{121/123}$Sb-NQR spectra has been shown to follow a Curie-Weiss temperature dependence~\cite{feng2023}. At $\sim$1.9~GPa, close to our $p^*$, the Weiss temperature is zero which has been regarded as evidence of a CDW QCP~\cite{feng2023}. Finally, a pressure-induced QCP around $p^*$ has been suggested by theoretical proposals~\cite{tazai2022,wang2022}. Therefore, taking into account high-pressure results from multiple probes, the pressure dependence of $I_{\rm c,sf}$ unambiguously confirms the occurrence of a quantum phase transition beneath the superconducting dome. Our work successfully demonstrates that the concept of quantum phase transition can be employed for optimizing the critical current of the kagome superconductor \CsVSb, and likely for many other systems as well. As an extension of the present work, the effect of chemical substitution on \CsVSb\ should be explored. For example, Sn-doping has been shown to give rise to a double $T_c$ dome~\cite{oey2022}. A careful $I_{\rm c,sf}$ study will further shed light on the interplay between the CDW phase and superconductivity.
\\

\noindent {\bf \large 3. Conclusion}\\
\noindent In summary, we have explored the possibility of enhancing the critical current of \CsVSb\ by driving the system through a quantum phase transition. We have investigated two \CsVSb\ thin flakes with thicknesses 
 of 225~nm and 180~nm via electrical transport measurements, and our temperature-pressure phase diagram reveals the rapid suppression of the CDW phase under pressure with $T_{\rm c}$ peaking near the CDW suppression pressure $p^* \approx$  20~kbar. Based on the temperature-pressure phase diagram, we have performed the self-field critical current measurements spanning a wide pressure range that includes $p^*$. The superconducting gap $\Delta_0$, extracted from a single $s$-wave gap fit, exhibits non-monotonic behaviour but closely tracks the pressure dependence of $T_{\rm c}$. Notably, the critical current exhibits a substantial enhancement of at least 10-fold near $p^*$ compared with that at ambient pressure. The fact that the critical current at the zero-temperature limit shows a peak at $p^*$ implies the presence of a quantum phase transition associated with the fading CDW phase in \CsVSb, which is consistent with other experimental data based on normal state properties~\cite{mu2021s,shan2022muon,zhang2024,Chen2021a,feng2023}. 
 Our findings thus highlight the possibly generic potential for increasing critical current density through the proximity of a quantum phase transition, paving the way for systematically enhancing the critical current in many other material families. 
 \\

\noindent {\bf \large 4. Experimental Section}\\
\noindent {\bf 4.1 Sample Synthesis, Characterization and Exfoliation}\\
\noindent High-quality single crystals of \CsVSb~were synthesized from Cs (ingot, 99.95 $\%$), V (powder, 99.9 $\%$) and Sb (shot, 99.9999 $\%$) using self-flux method similar to Refs.~\cite{Ortiz2019,Ortiz2020}. We sealed the raw materials with the molar ratio of Cs:V:Sb = 7:3:14 inside a pure-Ar-filled stainless steel jacket.
The as-grown single crystals were millimeter-sized shiny plates. X-ray diffraction (XRD) experiments of single crystals were conducted at room temperature using a Rigaku x-ray diffractometer
with CuK$\alpha$ radiation. The chemical compositions were determined by a JEOL JSM-7800F scanning electron microscope equipped with an Oxford energy-dispersive x-ray (EDX) spectrometer. Two \CsVSb\ thin flakes with thicknesses of 225~nm (S1) and 180~nm (S2), respectively, were exfoliated using the ``blue tape" (from Nitto Denko Co.) from bulk single crystals from the same batch. These flakes were then appended to the silicone elastomer polydimethylglyoxime (PDMS, Gelfilm from Gelpak) stamp, ready to be integrated into the high pressure diamond anvil cell.  
\\
\\
\noindent {\bf 4.2 Diamond Anvil Cells}\\
We employed the ``device-integrated diamond anvil cell" (DIDAC) technique~\cite{Xie2021,ku2022} to conduct the electrical transport measurements of both \CsVSb\ flakes under high pressure. Our pressure cells are equipped with two diamond anvils (Type IIas) with culet diameter of 800 $\mu$m and a bevel of 1000 $\mu$m. The schematic drawing of our diamond anvil cell is shown in Figure S1a (Supporting Information). We used the photolithography technique and physical vapour deposition (PVD) coating to attach a pattern of the six-probe Au microelectrodes on the culet of a diamond anvil. The alumina/Stycast-1266  mixture was placed between the stainless steel gasket and the microelectrodes as the insulating layer. The exfoliated \CsVSb\ thin flake was then transferred onto the center of the culet. After that, the \CsVSb\ thin flake was encapsulated by a h-BN thin film to protect the sample from oxidization. A top view of S1 positioned on the anvil culet can be seen in Figure S1b (Supporting Information). The thickness of the thin flakes was determined by a dual-beam focused ion beam system (Scios~2 DualBeam by Thermo Scientific). Finally, glycerin with a high purity of 99.5$\%$ was used as the pressure transmitting medium and we used the spectrum of ruby fluorescence at room temperature to calibrate the pressure on the sample.  
\\
\\
\noindent {\bf 4.3 Electrical Transport Measurements}\\
The electrical resistance and self-field critical current measurements were performed by a standard four-terminal configuration in the Physical Property Measurement System (PPMS) by Quantum Design and a dilution refrigerator by Bluefors. We used the six-probe Au microelectrodes mentioned above to build a robust electrical contact with the \CsVSb\ thin flake. The voltage-current ($V$-$I$) curves were measured using a Keithley 2182A nanovoltmeter in conjunction with a Keithley 6221 current source operating in the pulsed delta mode. The pulsed current had a duration of 11~ms, with a pulse repetition time of 1~s.
\\
\\

\begin{acknowledgments}
\noindent {\bf Acknowledgements}\\
The work was supported by Research Grants Council of Hong Kong (A-CUHK 402/19, CUHK 14301020, CUHK 14300722, CUHK 14302724), CUHK Direct Grant (4053577, 4053525), the National Natural Science Foundation of China (Grant No. 12174175, 12104384) and the Guangdong Basic and Applied Basic Research Foundation (Grant No. 2022B1515120014). \\
\end{acknowledgments}

\noindent {\bf  Conflict of Interest}\\
The authors declare no conflict of interest.\\

\noindent {\bf  Data Availability Statement}\\
The data that support the findings of this study are available from the corresponding author upon reasonable request.\\


\providecommand{\noopsort}[1]{}\providecommand{\singleletter}[1]{#1}%

\end{document}